\documentclass[12pt]{iopart}
\usepackage{epsfig}
\usepackage{graphicx}
\usepackage{dcolumn}
\usepackage{iopams}
\usepackage{epstopdf}

\expandafter\let\csname equation*\endcsname\relax 
\expandafter\let\csname endequation*\endcsname\relax 

\usepackage{amsmath}
\usepackage[english]{babel}
\usepackage{hyperref}
\usepackage{bm}

\begin{document}

\title[Fermionization in few-fermion systems]{Fermionization of two-component few-fermion systems in a
  one-dimensional harmonic trap} 

\author{E J Lindgren$^{1}$, J Rotureau$^{1}$, C Forss{\'e}n$^{1}$, A G Volosniev$^{2}$ and N T Zinner$^{2}$}
\address{$^{1}$ Department of Fundamental Physics, Chalmers University of
  Technology, SE-412 96 G{\"o}teborg, Sweden \\
  $^{2}$ Department of Physics and Astronomy, Aarhus University, DK-8000 Aarhus C, Denmark}

\date{\today}

\begin{abstract}
  The nature of strongly interacting Fermi gases and magnetism is one of
  the most important and studied topics in condensed-matter
  physics. Still, there are many open questions. A central issue is
  under what circumstances strong short-range repulsive interactions are
  enough to drive magnetic correlations. Recent progress in the field of
  cold atomic gases allows to address this question in very clean
  systems where both particle numbers, interactions and dimensionality
  can be tuned. Here we study fermionic few-body systems in a one
  dimensional harmonic trap using a new rapidly converging
  effective-interaction technique, plus a novel analytical
  approach. This allows us to calculate the properties of a single
  spin-down atom interacting with a number of spin-up particles, a case
  of much recent experimental interest. Our findings indicate that, in
  the strongly interacting limit, spin-up and spin-down particles want
  to separate in the trap, which we interpret as a microscopic precursor
  of one-dimensional ferromagnetism in imbalanced systems. Our
  predictions are directly addressable in current experiments on
  ultracold atomic few-body systems.
\end{abstract}

\maketitle

\section{Introduction}
Few-fermion systems are the building blocks of matter. Atoms and nuclei
are well-known examples, but also systems such as quantum dots,
superconducting grains, and other nanoscale structures are of great
interest. The key to understanding such structures is first and
foremost the relation between the discrete level structure, due to the
finite size, and the strength and nature of interparticle interactions. An
exciting recent development in atomic physics is the experimental
realization of few-body Fermi systems with ultracold atoms
\cite{serwane2011,zurn2012}.
These setups are extremely versatile as the
potential that traps the atoms can produce lattices and/or
low-dimensional geometries \cite{bloch2008}, and the atomic interaction
strength may be tuned via the use of Feshbach resonances
\cite{chin2010}. The spin-$1/2$ nature of electrons or nucleons is
addressable by populating two hyperfine states in the atoms and we thus
have a direct mapping from the atomic setup to ordinary matter. We will
refer to these two components as spin up and spin down.

A seminal contribution of ultracold atomic gas research is the
realization of strongly interacting quantum gases
\cite{paredes2004,kinoshita2004,kinoshita2005,haller2009} using
confinement-induced resonances \cite{olshanii1998}. The Tonks-Girardeau
(TG) gas \cite{tonks1936,girardeau1960,lieb1963} of strongly repulsive
bosons that displays fermionic behavior \cite{deuret2008,jonas2009} is one such example
\cite{paredes2004,kinoshita2004,kinoshita2005}.  The so-called {\it
  super}-TG (sTG) limit of very strong attractive interactions has also
been addressed both theoretically
\cite{astra2004,astra2005,batchelor2005,tempfli2009,
  girardeau2010a,girardeau2011,valiente2012,brouzos2012,girardeau2012} 
and experimentally \cite{haller2009}. Most recently, the TG and sTG
states have been explored in fermionic systems
\cite{guan2009,girardeau2010b,chen2010,guan2010,brouzos2013}. While the two-body
system in a harmonic trap has a well-known exact solution for any
interaction strength, known as the Busch model~\cite{busch1998},
two-component fermionic few-body 
systems with more than two particles have not been solved 
exactly. Although a number of numerical studies have been performed 
(see discussion below), many questions still remain related to the 
main difficulty in the handling of very strong interactions in the vicinity
of the fermionization limit.

\begin{figure}[ht!]
\centering
\includegraphics[scale=0.60]{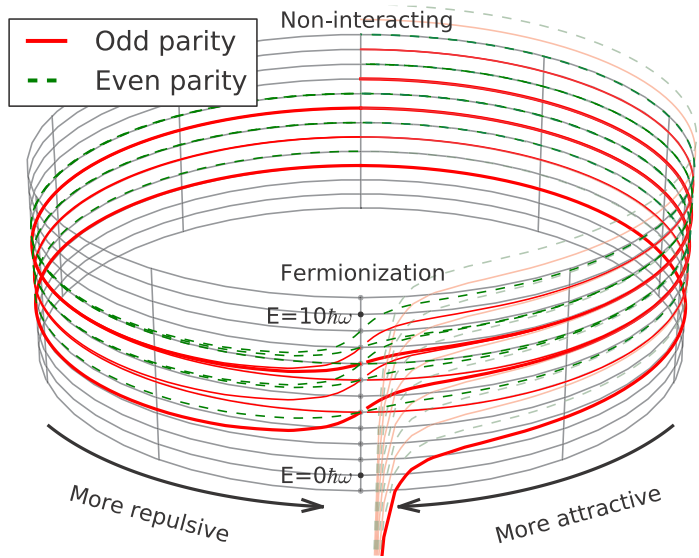}
\caption{Energy spectrum for a three-body system ($N_\uparrow=2,
  N_\downarrow=1$) in a harmonic trap as a function of the strength,
  $-1/g$, of the zero-range interaction between up and down components.
  The cylinder plot highlights the connections of the different states
  in the non-interacting ($g=0$) and strongly interacting ($|g|=\infty$)
  limits; on the back and the front of the cylinder, respectively. The
  states are ordered according to their spatial parity (odd or
  even). See also Fig.~\ref{fig-e}.}
\label{fig1}
\end{figure}

In this article we face this challenge and consider the experimentally
accessible situation of a harmonically trapped few-body system in one
dimension with $N_\uparrow$ spin up and $N_\downarrow$ spin down
fermions for the imbalanced case where $N_\uparrow> N_\downarrow=1$.
Zero-range interactions of strength $g$ are employed between different
spin components, while the identical spin particles remain
non-interacting by the Pauli exclusion principle. This is a few-body
analog of the fermionic polaron, which is currently under intense study
\cite{schiro2009,kohstall2012,koschorreck2012}.  We solve the few-body
problem for various interaction strengths using a numerical technique
inspired by developments in nuclear physics
\cite{jimmy2013,lisetskiy2008}. 
This method is exact for any interaction strength in the limit of infinite model space; it
yields both the energy spectrum and energy eigenstate wave functions and indeed shows excellent
convergence properties.
In addition, we present an exact solution for the
$N_\uparrow=2$ case in the fermionization limit of infinite interaction
strength, using an analytical model. The classic work of McGuire
\cite{mcguire1965,mcguire1966} solved the untrapped case with periodic 
boundary conditions for arbitrary
$N_{\uparrow}$ and $g$, but the trapped case has only been solved exactly
for $N_\uparrow=1$ \cite{busch1998} and for resonant interaction in three dimensions
\cite{werner2006}. 

\section{Results}
We are interested in obtaining numerically exact eigensolutions for the
system of trapped atoms with arbitrary strong, zero-range interaction between
different spin components. We use harmonic-oscillator (HO) units, in
which the Hamiltonian is 
\begin{align}\label{ham}
H=\sum_{i\sigma}\left(\frac{k_{i\sigma}^2}{2}+\frac{x_{i\sigma}^2}{2}\right)
+g\sum_{i\sigma,j\tilde\sigma,i>j}\delta(x_{i\sigma}-x_{j\tilde\sigma}),
\end{align}
where lengths are in units of the oscillator length $b = \sqrt{\hbar / m
  \omega}$, energies in units of the trap oscillator energy $\hbar
\omega$, $\sigma=\pm$, and $\tilde\sigma=-\sigma$.  The
interaction strength, $g$, becomes dimensionless in units of $\hbar
\omega b$.  These units are used for all quantities, unless we
explicitly state otherwise. The Hamiltonian is parity invariant and one
can classify states as either even or odd under $x\to-x$.
We concentrate mostly on the first non-trivial case beyond the
two-fermion system, which is $N_\uparrow=2$, $N_\downarrow=1$ (denoted
2+1).  The many-body problem is solved using an effective-interaction
approach that uses the exact analytical solution of the two-body problem
as input. As we discuss below in \ref{conv}, this speeds up the
convergence tremendously and allows us to obtain very accurate results
for mesoscopic samples with particle numbers of order ten.  We stress
that our approach is far superior to exact diagonalization with the bare
zero-range interaction, which has a very slow convergence (see \ref{conv}). 
The numerical method used in this work therefore represents a
significant advance in the description of strongly interacting,
finite-size quantum systems.

\begin{figure}[ht!]
\centering
\includegraphics[scale=0.6]{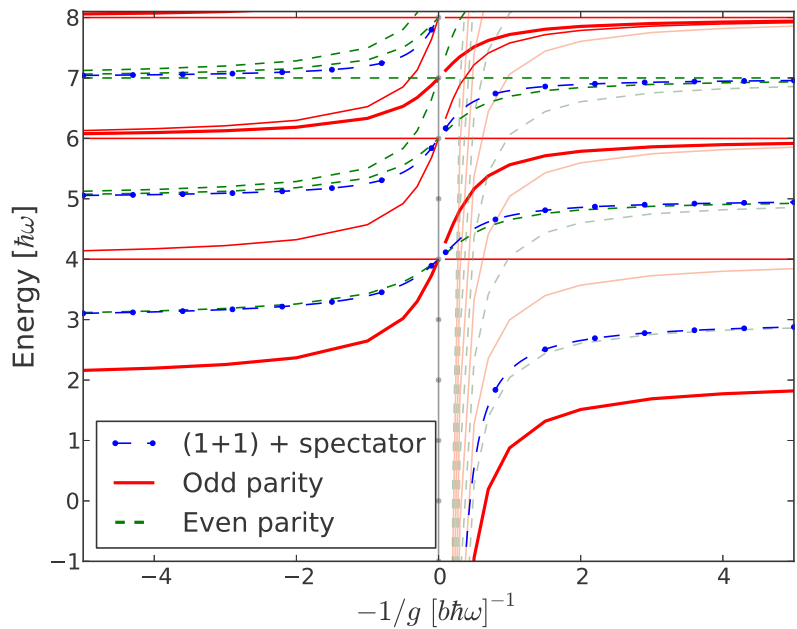}
\caption{Energy spectrum for the 2+1 system (excluding the center-of-mass
	contribution). 
	Diverging states on the
  attractive side are dimmed. The Busch-model for a 1+1 system
plus the energy of a spectator particle has been added for comparison.}
\label{fig-e}
\end{figure}

In Fig.~\ref{fig1} we show the numerically obtained energy spectrum for
the 2+1 system; plotted on a cylinder where the fermionization limit
$|g|\to\infty$ is on the front while weak-coupling $|g|\to 0$ is on the
back. We plot the total intrinsic energy,
i.e. the total energy minus $\hbar\omega/2$ from the center of mass
motion in the trap.
This way of plotting the spectrum emphasizes the spectral flow and
the connection to the Zeldovich rearrangement effect
\cite{farrell2012}. The first interesting feature is the ground state
behavior; it starts as a strongly bound dimer plus a particle for
$g\to-\infty$, wraps around the cylinder to the non-interacting limit,
$|g|\to0$, and then becomes energetically degenerate with two other
states at $g\to\infty$. 
Another representation of the 2+1 spectrum for odd and even parity
states is shown in Fig.~\ref{fig-e}. The horizontal lines correspond to
totally antisymmetric states, which are non-interacting in the case of
zero-range interactions. For example, the lowest such state, at $E = 4
\hbar\omega$, corresponds to having one particle in each of the three
lowest HO states. At $g=\infty$ it becomes degenerate with two
interacting states. Note also that we have many molecular branches close
to $|g|=\infty$ for $g<0$ (dimmed curves in Fig.~\ref{fig-e}).  Starting
from $g\to 0^+$ (far left in the figure) and following the odd ground
state we see that it makes a jump around $g=\infty$ before becoming
non-interacting at $g\to 0^-$ (far right). This is an analog of the
so-called repulsive branch for untrapped polarons
\cite{kohstall2012,koschorreck2012}.  Repulsive branch means that
excited states are pushed up on the attractive side of the
$|g|\to\infty$ resonance in constrast to the lower-lying molecular
branches that become strongly bound, as shown in Figs.~\ref{fig1} and
\ref{fig-e}. However, the jump endured by the odd and even states that
become degenerate at $g=\infty$ is quite different.  For comparison, we
plot the two-body Busch results shifted by the energy of a free
spectator particle (dashed blue line in Fig.~\ref{fig-e}), which turns
out to be almost identical to the even parity state at low energy. This
even parity state therefore has an atom-dimer structure, with almost no
interaction between atom and dimer. This has also been observed in
three-dimensional traps \cite{daily2010}.

\begin{figure}[ht!]
\centering
\includegraphics[scale=0.60]{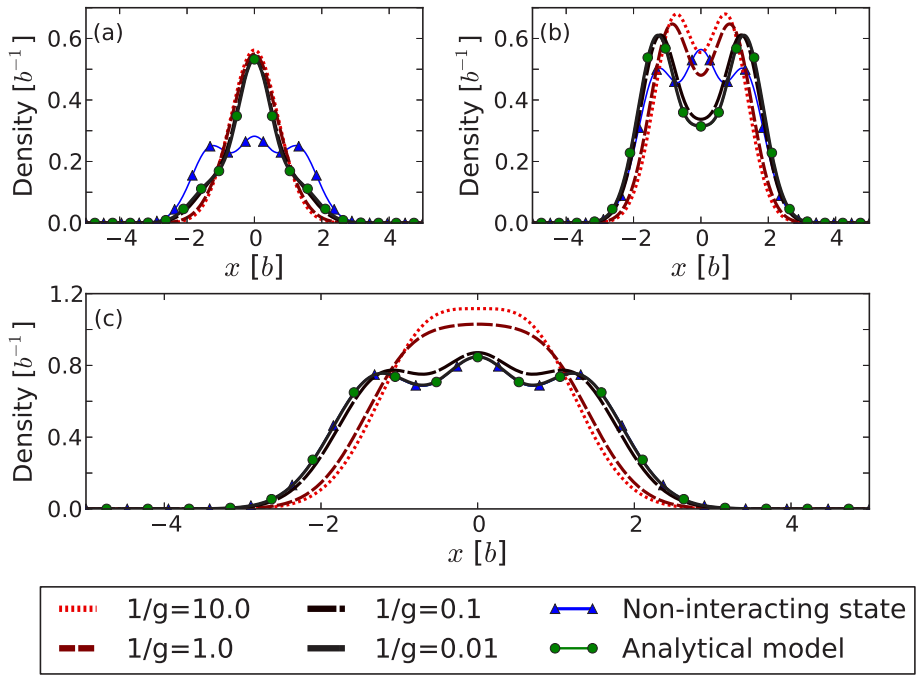}
\caption{Density distributions of the 2+1 ground state as a function of
  the (repulsive) interaction strength, $g$. Panels (a) and (b) show the
spin-separated density distributions for the impurity (spin-down) and
majority (spin-up) particles, respectively. Panel (c) corresponds to
the total density. The density of the non-interacting totally
antisymmetric state is plotted for comparison together with the
analytical model for the odd-parity ground state at fermionization.}
\label{fig2odd}
\end{figure}

A particularly interesting feature of the interacting states, as they
cross over to the attractive side of $|g|\to\infty$, 
is their density distributions that
we show in Fig.~\ref{fig2odd} for the odd ground state.  While they are
approaching the fermionization limit $|g|\to\infty$ for the total
density, the spin-resolved densities demonstrate a distinct separation
in the trap.  This we interpret as a precursor of ferromagnetic behavior
in a one-dimensional few-body context for imbalanced systems.  
In the vicinity of $|g|=\infty$, the ground, first excited, and
non-interacting states all have completely different spin-resolved
densities; the ground state has the impurity at the center and the first
excited state has the impurity at the edge while the non-interacting
state yields a three-hump profile independent of spin. This has all been 
verified using our analytical model (see \ref{anal}).
Since the states are
degenerate at $|g|=\infty$, this clearly demonstrates that the behavior
is not due to energetics but to different correlations in the wave
functions.  It also shows that the fermionization limit is very
different for two-component fermions as compared to fermionization of
bosons.

\begin{figure}[ht!]
\centering
\includegraphics[scale=0.6]{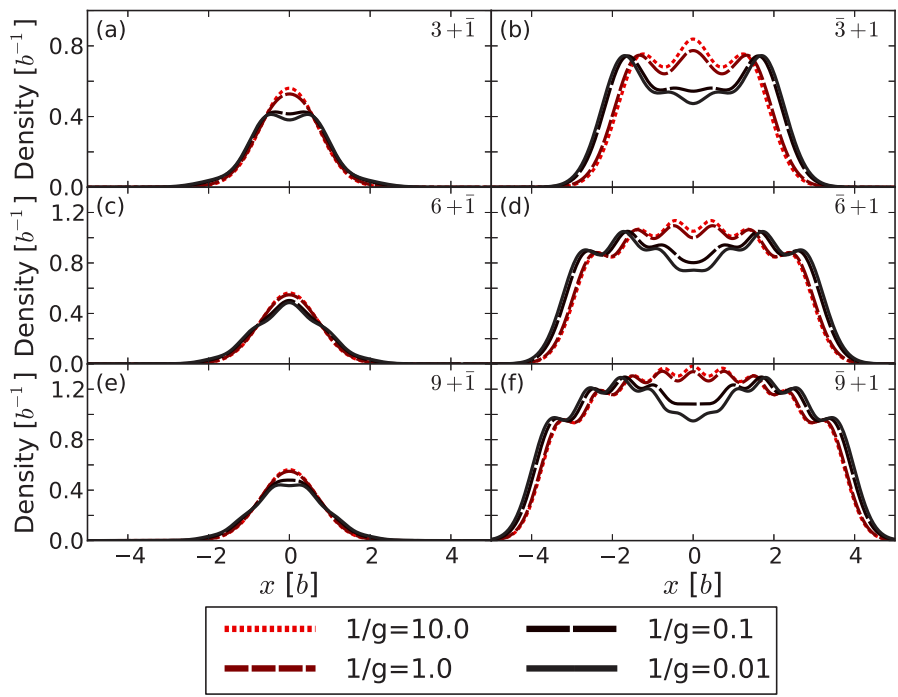}
\caption{Spin-resolved densities for the 3+1, 6+1, and 9+1 systems,
  cf. Fig.~\ref{fig2odd}. Panels (a), (c) and (e) show the distribution of
  the impurity particle, while panels (b), (d) and (f) show the majority
  density.}
\label{fig3}
\end{figure}

The approach that we have presented can be applied to larger systems. In
Fig.~\ref{fig3} we show the spin-resolved densities for the ground
states of the 3+1, 6+1, and 9+1 systems as a function of the (repulsive)
interaction strength. These spin densities show the same spin
separation behavior in the limit of fermionization as the 2+1 case in
Fig.~\ref{fig2odd}. Our results imply that this is a more general
feature of one-dimensional two-component Fermi systems.
These general structures can be experimentally investigated by
tunneling \cite{zurn2012,rontani2012}.

\section{Discussion}
\begin{figure}[ht!]
\centering
\includegraphics[scale=0.60]{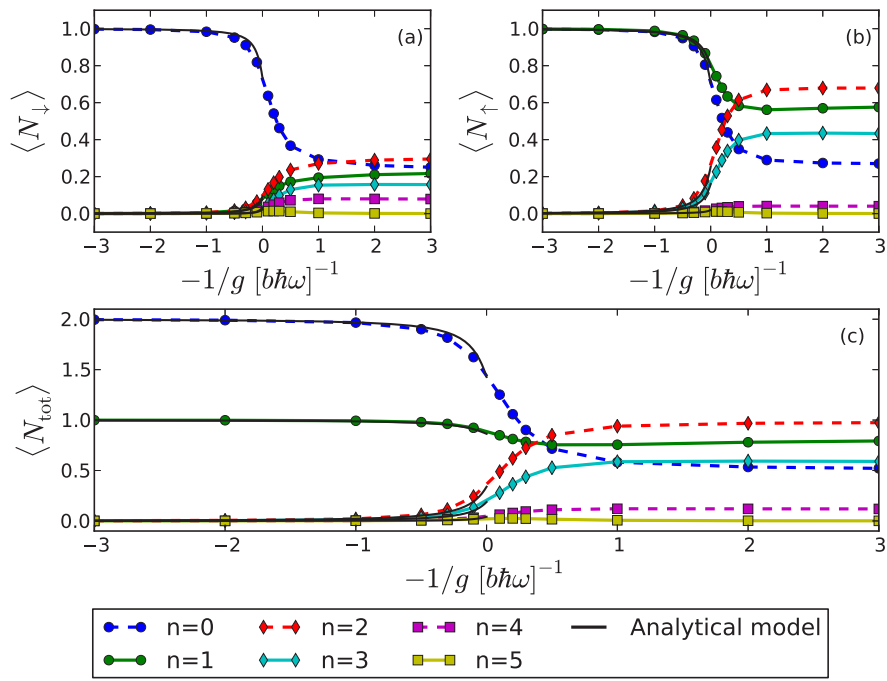}
\caption{Occupations numbers as a function of the interaction strength
  for the ground state of the 2+1 system. Panels (a) and (b) show the
  spin-separated occupation numbers, while (c) corresponds to the total
  occupation number.}
\label{fig:occ}
\end{figure}

Current experiments on few-fermion systems \cite{serwane2011,zurn2012}
can study the structures that we predict by performing tunneling
measurements that map out the occupancies of the few-body wave
function. By varying $g$ one can explore the structure on both sides of
the resonance \cite{zurn2012}. It is possible to go diabatically from
the repulsive ground state and onto the repulsive branch on the $g<0$
side since the overlap with the atom+dimer molecular branch is small.
It is thus possible to investigate a large part of the parameter
space. In Fig.~\ref{fig:occ} we show the occupancies of different
single-particle levels in the trap. Note how well our analytical model
reproduces the numerical results for $g>0$.  By selective ejection of
the minority particle it is possible to measure the majority occupation
number. A preliminary comparison to experimental data shows agreement
with our predictions \cite{jochim}.

Our numerical and analytical findings show that
around fermionization the two spins tend to separate in the trap. We
interpret this as a few-body precursor of Stoner ferromagnetism
\cite{stoner1933} in one dimensional imbalanced systems.  Ferromagnetism
is hotly pursued topic at the moment both in balanced and imbalanced Fermi 
systems
\cite{jo2009,cui2010,pekker2011,zhang2011,pietro2011,sanner2012,cui2012,pietro2013}.  
As discussed above, this separation of species should be directly
observable in current experiments
\cite{zurn2012,gerhard-thesis,sala2013}. Other methods have been
employed recently to similar systems to study energy spectra
\cite{harshman2012,gharashi2012} and also the density profiles
\cite{brouzos2013}. Here we have presented a complementary method that
converges extremely fast for multi-particle systems. We have also 
provided new analytical insights into the problem by obtaining a 
wave function for $g>0$ that becomes exact at $g\to \infty$ and 
reproduces both degeneracies, densities, and occupation numbers 
(see \ref{anal}). Lastly, we note that a recent paper by 
Gharashi and Blume \cite{gharashi2013} has studied some of the 
same systems using a different method. 
Our results for similar systems (2+1 and 3+1) 
are in agreement with reference~\cite{gharashi2013}, 
and both are in agreement with the exact solution 
for large repulsive interactions presented in 
reference~\cite{volosniev2013}. However, they do 
not agree with the results based on symmetry arguments 
presented in reference~\cite{guan2009}. The reasons 
that symmetry and group theoretical arguments and 
spin algebra cannot be used to determine the 
eigenstates for large but finite interaction strengths 
are discussed in the appendix of 
reference~\cite{volosniev2013} and some explicit 
examples are given in the supplementary material 
of reference~\cite{gharashi2013}.

Magnetism is often considered a bulk property of a system while 
magnetic correlations such as super-exchange, etc., are typically
discussed in the context of just a few particles. Many studies
of magnetism in one dimension are conducted in a flat potential
and employing periodic boundary conditions starting from the 
work of McGuire \cite{mcguire1965,mcguire1966}. Dispensing of 
the periodic boundaries (which are not suitable for the 
few-body systems studied here), we may consider how our results would
change if we had replaced the harmonic trap with a hard-wall 
box potential. Since the degeneracy in the strongly interacting
limit is a result of short-range correlations in the wave 
function (nodal structures), we do not expect anything to change 
there. However, since the approach to the strongly interacting
regime certainly depends on the single-particle wave functions 
(as clearly seen in the analytical model presented here)
the spectrum will change quantitatively under the constraint
that the degeneracies at infinite coupling strength are preserved.

The effective-interaction approach used in this work is key to the
quality of our numerical results and to our conclusions. In the
construction of these effective interactions we benefit from having
access to the exact two-body solutions. However, we stress that, using
numerical two-body solutions, this approach can be generalized to study
many-body systems in higher dimensions, with finite-range interactions,
and in any trapping potential.

\ack
Discussions with D. Blume, S.~E. Gharashi, A.~S. Jensen, D.~V. Fedorov, 
and G.~M. Bruun are highly appreciated. We are grateful to the 
group of S. Jochim in Heidelberg for discussions and for sharing unpublished
data with us.
The research leading to these results has
received funding from the 
European Research Council under the European Community's Seventh
Framework Programme (FP7/2007-2013) / ERC grant agreement
no.~240603. This work was also supported by the Swedish Research
Council (dnr.~2007-4078).

\appendix
\section{Effective interaction approach}\label{conv}
We solve numerically the many-body Schr\"odinger equation with the
Hamiltonian~\eqref{ham} in a finite basis constructed from the HO
single-particle states $|n\rangle$. Each many-body basis state is
written as $|n_{1} \ldots n_{N\uparrow}\rangle
\otimes|n_{\downarrow}\rangle$ {\it i.e.}  a product of a HO
antisymmetrized state of the $N_\uparrow$ spin up particles and a HO
single particle state for the spin down particle. The corresponding HO
energy is $(n_1+\ldots+n_{N\uparrow}+n_\downarrow+\frac{N}{2})$ where
$N=N_\uparrow+N_\downarrow$ is the total number of particles.  The
model space truncation is defined by a total upper limit, i.e.
$n_1+\ldots+n_{N\uparrow}+n_\downarrow\leq n_\mathrm{tot}$. Since we
are only interested in the intrinsic dynamics of states, a Lawson
projection term~\cite{gloeckner1974} is used to push away many-body
solutions corresponding to excitations of the center of mass motion.

Instead of the bare zero-range interaction in (\ref{ham}), we consider
an effective two-body interaction in order to speed up the convergence
of the eigenenergies with respect to the size of the many-body
basis. The effective potential $V^\mathrm{eff}_{P}$ is constructed in a
truncated two-body space $P$, defined as the set of two-body relative HO
states whose radial quantum number are smaller or equal to a cutoff
$n_{P}$. The effective force $V^\mathrm{eff}_{P}$ is designed such that
its solutions correspond to exact solutions given by the Busch formula
\cite{busch1998}. Using a unitary transformation, we construct a
two-body effective Hamiltonian $H^{\rm{eff}}_{P}$ as \cite {jimmy2013}
\begin{eqnarray}
H^{\rm{eff}}_{P}=\frac{U_{PP}^{\dagger}}{\sqrt{(U_{PP}^{\dagger}U_{PP})}}
E^{(2)}_{PP}
\frac{U_{PP}}{\sqrt{(U_{PP}^{\dagger}U_{PP})}},
\label{hami_eff} 
\end{eqnarray} 
where $E^{(2)}_{PP}$ is the diagonal matrix formed by the $n_P+1$ lowest
exact energies given by the Busch formula, and $U_{PP}$ is the matrix
whose rows are formed by the corresponding eigenvectors projected on
$P$. The effective interaction $V^\mathrm{eff}_{P}$ is obtained from
$H^{\rm{eff}}_{P}$ by subtracting the HO potential.  For each cutoff
$n_{P}$, we diagonalize the many-body Schr\"odinger equation with
$V^\mathrm{eff}_{P}$ and increase $n_\mathrm{tot}$ until convergence of
the many-body energies is reached \cite{jimmy2013}. We find that
$n_\mathrm{tot} = n_P + 2$ is sufficient to capture the properties of
the effective interaction. With this choice, we can then study the
energy convergence as a function of $n_\mathrm{tot}$. By construction,
this unitary transformation approach will reproduce exact bare
Hamiltonian results (both energy spectrum and wave functions) in the
$n_P \to \infty$ limit.

\begin{figure}[ht!]
\centering
\includegraphics[scale=0.6]{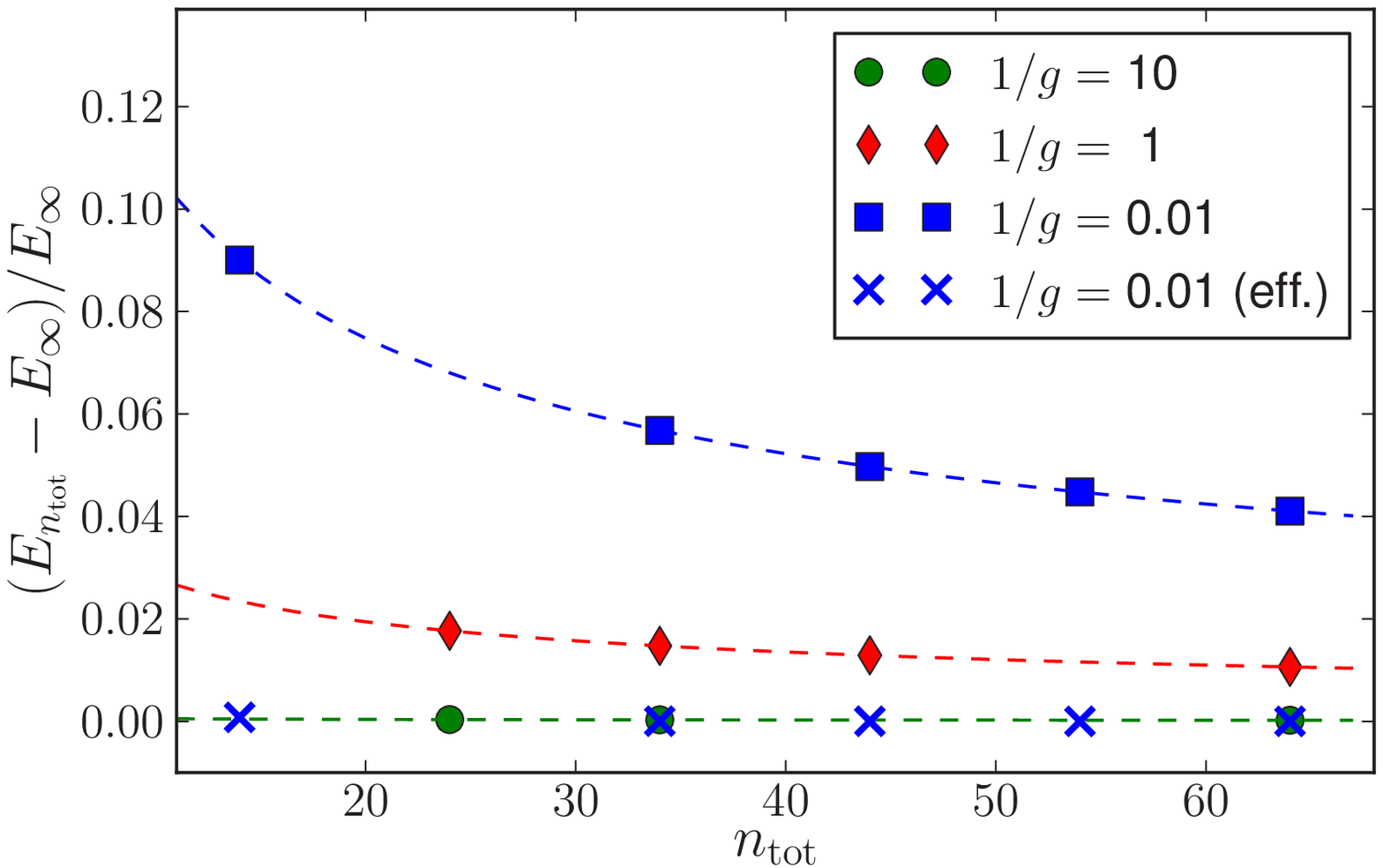}
\caption{Relative error $(E_{n_\mathrm{tot}}-E_{\infty})/E_{\infty}$ in the ground state energy of the $2+1$ system
  as a function of the many-body model-space truncation
  $n_\mathrm{tot}$. We show results for several bare interaction
  strengths (represented by circles, diamonds, and squares), and the
  effective-interaction results for the strongest case, $1/g = 0.01$,
  (crosses). In this latter case, $n_\mathrm{tot} = n_P+2$, where $n_P$
  is the truncation of the two-body space.%
\label{fig-lz}}
\end{figure}
The excellent convergence property of our method is demonstrated in
Fig. \ref{fig-lz} that shows the relative error
$(E_{n_\mathrm{tot}}-E_{\infty})/E_{\infty}$ in the ground state energy
of the $2+1$ system as a function of the size of the model space
$n_\mathrm{tot}$. We show results using the bare interaction for several
different strengths, and one sequence of results obtained with the
effective interaction $V^\mathrm{eff}_{P}$ for the strongest case. For
each interaction strength, we define $E_\infty$ as the converged result
obtained with $V^\mathrm{eff}_{P}$. Fast convergence will then be
characterized by the relative difference being close to zero already for
small $n_\mathrm{tot}$. As expected, with the bare interaction, the
convergence with $n_\mathrm{tot}$ is much slower for the strong
interactions: for $1/g=1$ the relative error is $\sim$ 1\% for
$n_\mathrm{tot}=64$, whereas for $1/g=0.01$, the relative error is still
$\sim$ 4\% for the same model space.  On the other hand, for $1/g=0.01$,
the result obtained with the effective interaction is within 0.01\% of
the fully converged value already at $n_\mathrm{tot}=14$. Dashed lines
correspond to a fit to the bare interaction results using the functional
form: $E_{n_\mathrm{tot}} = E_\infty+ c n_\mathrm{tot}^{-\lambda}$, with
$E_\infty, c, \lambda$ as free fit parameters. %

\section{Analytical model}\label{anal}
We now present an analytical approach that captures the behavior of the
wave function exactly at $g=\infty$.  We first define Jacobi
coordinates, $x=(x_{1\uparrow}-x_{2\uparrow})/\sqrt{2}$ and
$y=\sqrt{2/3}x_{3\downarrow}-(x_{1\uparrow}+x_{2\uparrow})/\sqrt{6}$,
and use these to obtain the spherical variables, $r=\sqrt{x^2+y^2}$ and
$\tan\phi=y/x$.  For $g=0$, the (unnormalized) eigenstates are
$r^{\mu}L_{\nu}^{\mu}(r^2)e^{-r^2/2}\cos(\mu\phi)$ and
$r^{\mu}L_{\nu}^{\mu}(r^2)e^{-r^2/2}\sin(\mu\phi)$, where $\mu$ and
$\nu$ are non-negative integers and $L_{\nu}^{\mu}(z)$ is the associated
Laguerre polynomial. The corresponding energies are $E=2\nu+\mu+1$.
We need only
consider the wave function in the interval $0<\phi<\pi/2$ since one can
use the Pauli principle and parity invariance to extend to
$0<\phi<2\pi$.
At $\phi=\pi/6$ opposite spins overlap. The full solution can thus be obtained by
matching the wave function and its derivative on the line $\phi=\pi/6$. We have
\begin{align}
&F_1(r,\pi/6)=F_2(r,\pi/6)\quad\textrm{and}\quad&\label{cond1}\\
&\frac{1}{2r^2}\frac{\partial F_{1}(r,\pi/6)}{\partial\phi}-\frac{1}{2r^2}\frac{\partial F_{2}(r,\pi/6)}{\partial\pi}=-\frac{g}{\sqrt{2}r}F_1(r,\pi/6),\label{cond2}&
\end{align}
where $F_1$ and $F_2$ are solutions for $0<\phi<\pi/6$ and
$\pi/6<\phi<\pi/2$ respectively. For $g\neq 0$ these equations are
complicated to solve, but by introducing an {\it ad hoc}
rescaled strength parameter $g_0=gr$ we decouple the equations and can
write $F_i(r,\phi)=R_i(r)\Psi_i(\phi)$ for $i=1,2$, where
$R_i(r)=r^{\mu}L_{\nu}^{\mu}(r^2)e^{-r^2/2}$. This rescaled 
model becomes exact when $|g|\to\infty$ \cite{zinner2013}.
The eigenfunctions and
eigenvalue equations can now be obtained by using the free angular
solutions $A\cos(\mu\phi)+B\sin(\mu\phi)$ and the conditions in
Eqs.~\eqref{cond1} and \eqref{cond2}.  The nature of the three-fold
degeneracy at fermionization seen in Fig.~\ref{fig1} and
Fig.~\ref{fig-e} comes from the odd and even solutions of these
equations, while the non-interacting has the 
structure $\cos(3\phi)$.
In the case of $\mu>0$, the angular wavefunction for
odd parity become $\Psi_1=N_o\sin(\mu\pi/3)\cos(\mu\phi)$ and
$\Psi_2=N_o\sin(\mu(\pi/2-\phi))\cos(\mu\pi/6)$, and for even parity
$\Psi_1=N_e\sin(\mu\pi/3)\sin(\mu\phi)$ and
$\Psi_2=N_e\sin(\mu(\pi/2-\phi))\sin(\mu\pi/6)$, while the energies
can be obtained from the algebraic equations for odd, 
$\mu\cos(\mu\pi/2)+g_0\sqrt{2}\cos(\mu\pi/6)\sin(\mu\pi/3)=0$, and even
$\mu \sin(\mu\pi/2)+g_0\sqrt{2}\sin(\mu\pi/6)\sin(\mu\pi/3)=0$
solutions.  $N_o$ and $N_e$, are normalization factors. The important
point is that $N_o \sin(\mu \pi/3)$ and $N_e\sin(\mu \pi/3)$ are
non-zero in the limit $\mu\to 3$ and thus the wave functions are
non-zero.

\end{document}